\documentclass[aps,pra,notitlepage,twocolumn,superscriptaddress,showpacs]{revtex4-1}

\usepackage{color}
\usepackage{graphicx}
\usepackage{amsmath}
\usepackage{hyperref}

\begin{document}

\title{Optical coherences and wavelength mismatch in ladder systems}

\author{A. Urvoy}
\affiliation{5. Physikalisches Institut, Universit\"{a}t Stuttgart, Pfaffenwaldring 57, 70550 Stuttgart, Germany}

\author{C. Carr}
\affiliation{Joint Quantum Center (JQC) Durham-Newcastle, Department of Physics, Durham University, Rochester Building, South Road, Durham DH1 3LE, UK}

\author{R. Ritter}
\affiliation{5. Physikalisches Institut, Universit\"{a}t Stuttgart, Pfaffenwaldring 57, 70550 Stuttgart, Germany}
\affiliation{Joint Quantum Center (JQC) Durham-Newcastle, Department of Physics, Durham University, Rochester Building, South Road, Durham DH1 3LE, UK}

\author{C. S. Adams}
\affiliation{Joint Quantum Center (JQC) Durham-Newcastle, Department of Physics, Durham University, Rochester Building, South Road, Durham DH1 3LE, UK}

\author{K. J. Weatherill}
\affiliation{Joint Quantum Center (JQC) Durham-Newcastle, Department of Physics, Durham University, Rochester Building, South Road, Durham DH1 3LE, UK}

\author{R. L\"{o}w}
\affiliation{5. Physikalisches Institut, Universit\"{a}t Stuttgart, Pfaffenwaldring 57, 70550 Stuttgart, Germany}


\begin{abstract}
We investigate experimentally and theoretically the coherent and incoherent processes in open 3-level ladder systems in room temperature gases and identify in which regime electromagnetically induced transparency (EIT) occurs. The peculiarity of this work lies in the unusual situation where the wavelength of the probe field is shorter than that of the coupling field. The nature of the observed spectral features depends considerably on the total response of different velocity classes, the varying Doppler shifts for bichromatic excitation fields, on optical pumping to additional electronic states and transit time effects. All these ingredients can be absorbed in a model based on optical Bloch equations with only five electronic states.
\end{abstract}


\maketitle

\section{Introduction}

Since its first demonstration over two decades ago, electromagnetically induced transparency (EIT) \cite{Boller1991} has developed into one of the most popular spectroscopic methods in atomic physics \cite{Fleishauer} and has also been applied to molecular \cite{Benabid2005} and solid state physics \cite{Liu2009, Safavi-Naeini2011}. EIT can allow the observation of sub-natural linewidth spectral features, even in room temperature thermal vapours. Although precision spectroscopy is best suited to ultracold gases, for example in optical lattice clocks \cite{Nicholson12}, the apparatus required for such experiments is somewhat bulky and inconvenient when it comes to real world applications. In addition, many atomic species, and most molecules, cannot be laser-cooled. Consequently, devices based on thermal vapours are desirable. In this respect, the commercially available chip scale atom clock SA.45S from Symmetricom \cite{Symmetricom} presents a technological paradigm. The portfolio of fully integrated atomic devices could be expanded to frequency references \cite{Haensch71,Bell,Abel}, Faraday rotators \cite{Siddons09}, slowing and storing light \cite{Phillips01}, optical filters \cite{Menders91}, and many new optical functionalities when combined with micro-optical components \cite{Stern12}. Vapour cell experiments have also recently allowed the demonstration of the cooperative Lamb-shift \cite{Keaveney2012} and extreme dispersion \cite{Keaveney2012a}.
A special class within this field is added by the use of highly excited Rydberg states, whose large polarizability and strong interactions allow new features to thermal vapour cell experiments such as e.g. electric field sensors \cite{Daschner2012}, microwave detectors \cite{Sedlacek2012}, electro-optic modulation \cite{Mohapatra2008} and possibly as a single photon source \cite{Dudin2012, Baluktsian2012}. Most of the examples listed above rely on two color excitation schemes, where the manifold of electronic levels allow for various excitation pathways. 

Although the majority of work on EIT has used $\Lambda$-systems which take advantage of two stable ground states, it is also possible to extend EIT to a ladder system under certain conditions. Since EIT relies on the coherent superposition of two excitation pathways, it is favourable that the upper state lives sufficiently long, as given for high-lying Rydberg states \cite{Mohapatra}, but it works also with short-lived states \cite{Gea-Banacloche1995}. The exaggerated properties of Rydberg states \cite{LoewReview} opens new and exciting opportunities for non-linear quantum optics \cite{PritchardReview} particularly at the single-photon level \cite{Dudin2012,Peyronel,Maxwell} 

\begin{figure}[b]
    \includegraphics[width=3in]{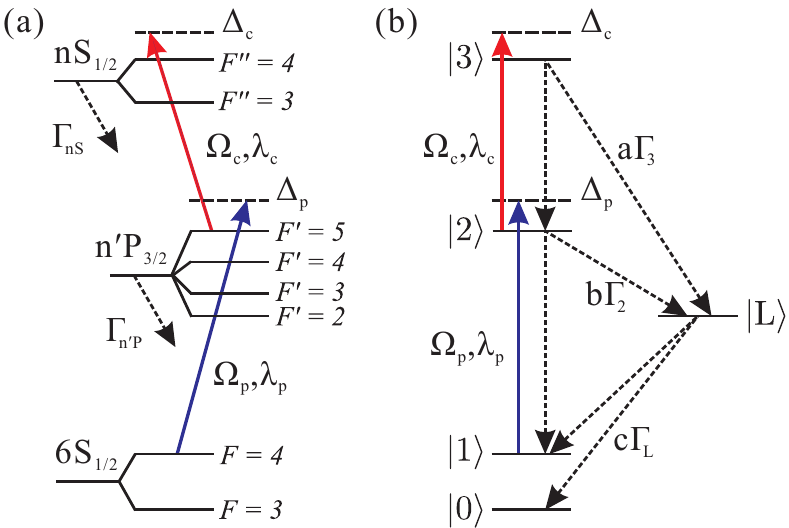}
    \caption{(a) Energy level scheme in caesium used experimentally. ${\rm n^{\prime}P, nS}$ are ${\rm 6P, 7S}$ and ${\rm 7P, 32S}$ for Section~\ref{section:Durham} and ~\ref{section:Stuttgart} respectively. ${\rm \Omega_{\rm p,c}}$, ${\rm \Delta_{\rm p,c}}$ and ${\rm \lambda_{\rm p,c}}$ are the Rabi frequencies, detunings and wavelengths of the probe and coupling laser field respectively. ${\rm \Gamma_{n^{\prime}P}}$ and ${\rm \Gamma_{nS}}$ are the decay rates of the excited states. (b) Energy level scheme used in the theoretical model. ${\rm \Gamma}_i $ is the decay rate of the state $| i \rangle $. a, b and c are the branching ratios from the three upper states ${\rm | 3 \rangle}$, ${\rm | 2 \rangle} $ and the loss state ${\rm | L \rangle}$ respectively. }
    \label{fig:fig1}
\end{figure}

\begin{figure*}[t]
    \includegraphics[width=6in]{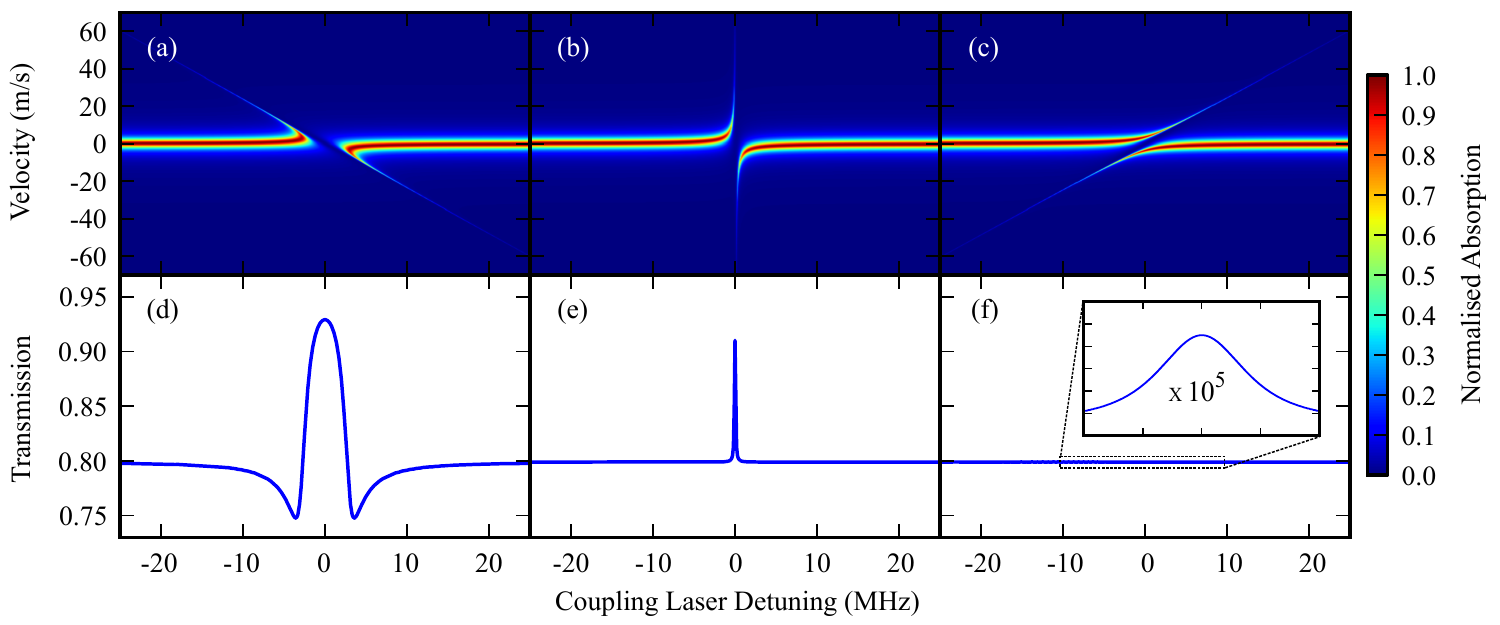}
    \caption{Wavelength dependence of the EIT. Top: Absorption coefficient per velocity class for (a) $\lambda_{\rm p}=\frac{4}{3} \lambda_{\rm c}$ ($\lambda_{\rm p}>\lambda_{\rm c}$) (b) $\lambda_{\rm p}=\lambda_{\rm c}$ and (c) $\lambda_{\rm p}=\frac{2}{3} \lambda_{\rm c}$ ($\lambda_{\rm p}<\lambda_{\rm c}$). Bottom (d)-(f): Doppler-averaged probe laser transmission signal corresponding to (a)-(c). Simulation parameters: $\Omega_{\rm p}=2\pi\times 0.01$~MHz, $\Omega_{\rm c}=2\pi\times 5$~MHz, $\Gamma_{1}=2\pi\times 5$~MHz, $\Gamma_{2}=2\pi\times 0.1$~MHz, ${\rm T}=20^{\circ}$C and $2$~mm cell length.}
    \label{fig:fig2}
\end{figure*}

The main focus of this work is to consider the wavelength ratio of the two excitation lasers used in ladder EIT \cite{Shepherd1996} and the effects of open decay channels in multi-level atoms. To obtain a detailed understanding of all possible dependencies, we perform spectroscopy on two different ladder schemes in caesium, which mainly differ in the lifetime of the upper state. When the upper excited state is short-lived, we find a competition between two-photon absorption and optical pumping which depends on the atom-laser interaction time. When the excited state is long-lived, we find that optical pumping effects are negligible and EIT remains observable as long as the decoherence mechanisms are minimized. 

In an EIT-ladder scheme, where the probe beam couples the two lower states and the coupling beam the two upper states, we will sometimes encounter a situation where the probe wavelength $\lambda_{\rm p}$ is smaller than the coupling wavelength $\lambda_{\rm c}$ ($\lambda_{\rm p}<\lambda_{\rm c}$) \cite{Carr2011, Carr2012a, Moon12}, what we call an ``inverted'' excitation scheme, since most published work on EIT in ladder systems employs light fields $\lambda_{\rm p}>\lambda_{\rm c}$. Such excitation schemes sometimes present a technological advantage as e.g. in the case of Rydberg states, where the small dipole matrix element can be matched by powerful and convenient infrared lasers. 

Let us first consider a simple 3-level ladder system, as shown in Figure~\ref{fig:fig1}~(b) with only the levels ${\rm | 1 \rangle}$, ${\rm | 2 \rangle} $ and ${\rm | 3 \rangle}$, ${\rm | 1 \rangle} \rightarrow {\rm | 2 \rangle} $ being the probe transition and ${\rm | 2 \rangle} \rightarrow {\rm | 3 \rangle} $ the coupling transition. A numerical solution of the optical Bloch equations for such a system, including the Doppler broadening, is depicted in Figure~\ref{fig:fig2} for three different wavelength ratios. In the upper panel, the absorption of the probe laser is shown as a function of atomic velocity and coupling laser detuning for the wavelength ratios (a) $\lambda_{\rm p}>\lambda_{\rm c}$ (b) $\lambda_{\rm p}=\lambda_{\rm c}$ and (c) $\lambda_{\rm p}<\lambda_{\rm c}$. In the lower panel, the Doppler averaged probe transmission signals, obtained by summing over the Gaussian-weighted signals for each atomic velocity class, are shown. In the case that $\lambda_{\rm p}<\lambda_{\rm c}$, the EIT feature is significantly smaller. This occurs because the one-photon Doppler shift $\Delta_{\rm 1ph}=-k_{p}v$ has the same sign as the two-photon Doppler shift $\Delta_{\rm 2ph}=-(k_{p}-k_{c})v$. As a result, the transparency window as a function of detuning no longer exists and adding the contribution from each velocity class strongly reduces the probe transmission peak compared to the non-inverted case. 

For real atoms the transitions in the three-level ladder system are not necessarily closed. Decay to other levels outside the three-level system will reduce the coherence time and by this the visibility of the EIT transparency window, and also introduce optical pumping effects. For instance when the upper transition is no longer closed, atoms can decay to the lower hyperfine level of the ground state via alternative decay channels. This process is known as Double Resonance Optical Pumping (DROP) \cite{Moon2004} and is a time-dependent transfer of atoms between the hyperfine levels of the ground state via a two-photon excitation process. Extensive studies of the interplay between EIT and DROP have been performed in a few non-inverted 3-level ladder system with theoretical models similar to that presented below (\cite{Hayashi10, Moon11, Noh11}).

In this paper we demonstrate experimentally and theoretically that in an ``inverted''-wavelength system, the interplay between these coherent and incoherent processes is strongly dependent upon the fine balance between state lifetimes, atom-laser interaction time, defined by the size of the laser beams and the velocities of the atoms, and the laser linewidths. The interplay of all these conditions leads to significant modifications of the optical signal observed. 

The experimental three-level ladder system in caesium that we use is shown schematically in Figure~\ref{fig:fig1}~(a). The ground state 6S$_{1/2}$ is coupled to the intermediate state n$^{\prime}$P$_{3/2}$ by the probe laser with Rabi frequency $\Omega_{\rm p}$ and wavelength $\lambda_{\rm p}$. This intermediate state is then coupled to the excited state nS$_{1/2}$ by the coupling laser with Rabi frequency $\Omega_{\rm c}$ and wavelength $\lambda_{\rm c}$.

\section{Model System}
\label{section:Model}

All levels and decays observed in the real level scheme (Figure~\ref{fig:fig1}~(a)) can be captured using a simplified model including 5 levels, as shown in Figure~\ref{fig:fig1}~(b). The actual states \mbox{$|6S_{1/2}, F=4 \rangle$}, \mbox{$|n^{\prime}P_{3/2}, F'=3,4,5 \rangle$} and \mbox{$|nS_{1/2}, F''=4 \rangle$} are represented by the states $|1\rangle$, $|2\rangle$ and $|3\rangle$ respectively. The uncoupled lower hyperfine level of the ground state 6S$_{1/2}, F=3$ is included as state $|0\rangle$. Finally, indirect decay from the intermediate and excited states to the ground states is taken into account through the addition of a loss state $|{\rm L}\rangle$. For instance, the decay \mbox{$|7S_{1/2}, F''=4 \rangle $ $\rightarrow$ $| 6P_{3/2}, F'=4 \rangle$ $\rightarrow$ $|6S_{1/2}, F=3 \rangle $} is represented by \mbox{$|3\rangle$ $\rightarrow$ $|L\rangle$ $\rightarrow$ $|0\rangle$} in the model. 

Using standard semi-classical methods \cite{cohen1992atom} we numerically solve the Liouville von Neumann master equation in steady-state for the density matrix $\rho$, which includes the populations and coherences of the five-level model. Due to thermal motion, the Doppler effect results in an effective detuning $\delta_{\rm Doppler}$ = $-\vec{k}\cdot \vec{v}$, where $k = 2\pi/\lambda$ is the wave vector of the laser field and $v$ the atomic velocity. As we use a co-linear beam configuration, this can be included in the model by averaging over the one-dimensional Gaussian velocity distribution. 

The finite interaction time of the atoms with the laser beam is included as a transit time decay. This is due to the component of the atomic motion orthogonal to the laser beam. Simply speaking, we re-distribute the atomic population of all states between the hyperfine levels of the ground state with a rate $\Gamma_{\rm tt}$ (\cite{Sagle1996}) with:
\begin{equation}
\Gamma_{\rm tt} = \frac{1}{a\sqrt{2\log2}}\sqrt{\frac{8k_{\rm B}T}{\pi m}}
\end{equation}
where $a$ is the $1/e^2$ beam waist, $T$ is the temperature and $m$ is the mass of the atom. 

Finally, we include laser-induced dephasing in the model. The finite linewidth of the laser can be modelled as a dephasing which acts on the off-diagonal elements of the density matrix $\rho$ \cite{Sultana1994, Gea-Banacloche1995}.

\section{Short-lived upper state}
\label{section:Durham}

\begin{figure}[t]
    \includegraphics[width=3.3in]{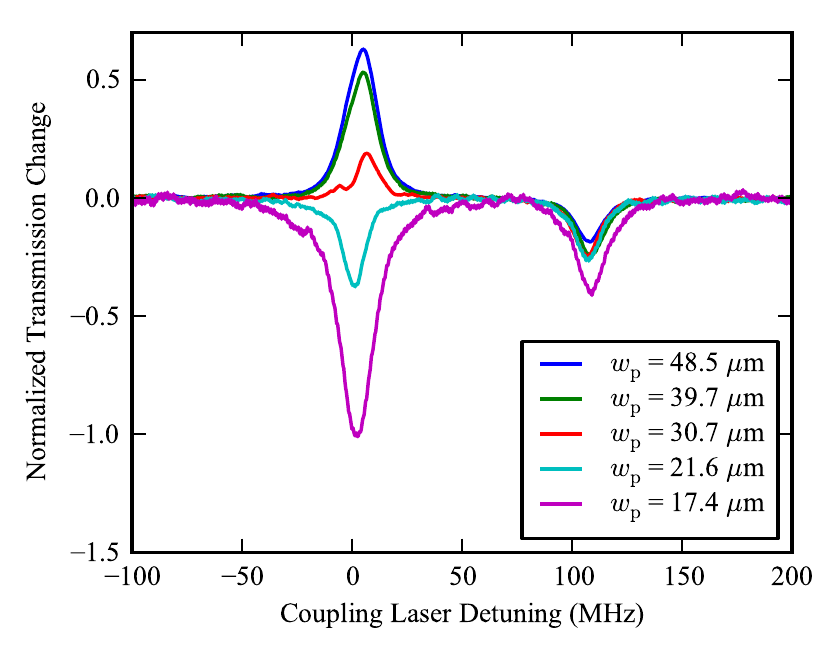}
    \caption{$6S-6P-7P$ experiment: Change in probe transmission as a function of probe beam waist $w_{\rm p}$ for the excitation with $\lambda_{\rm p} = 852$~nm and $\lambda_{\rm c} = 1470$~nm. Parameters: $\Omega_{\rm p}=2 \pi \times 8$~MHz, $\Omega_{\rm c}=2 \pi \times 22$~MHz and $w_{\rm c}=20$~$\mu$m.}
    \label{fig:fig3}
\end{figure}

First, we study a three-level system with ``inverted''-wavelengths where the upper level is a short-lived excited state. Experimentally, we realize this situation by tuning the the coupling laser to the $| 6P_{3/2} \rangle  \rightarrow | 7S_{1/2} \rangle $ transition. The $| 7S_{1/2} \rangle $ state has a natural linewidth of $\Gamma_{\rm 7S}=2\pi \times 3.3$~MHz comparable to the intermediate state linewidth of $\Gamma_{\rm 6P}=2\pi \times 5.2$~MHz. A circularly polarized 852 nm probe beam, stabilized to the $|6S_{1/2}, F=4 \rangle $ $\rightarrow$ $| 6P_{3/2}, F'=5 \rangle$ transition, passes through a caesium vapour cell. A counter-propagating circularly polarized 1470~nm coupling beam is scanned across the $| 6P_{3/2}, F'=5 \rangle $ $\rightarrow$ $| 7S_{1/2}, F''=4 \rangle $ transition. 

In order to investigate the atomic dynamics on different timescales, we use tightly focused beams in a 100~$\mu$m long vapour cell. The beam waist of the coupling beam is kept constant at $w_{\rm c}=20$~$\mu$m, and the beam waist of the probe beam $w_{\rm c}$ is varied from 17~$\mu$m to 49~$\mu$m. The beam sizes are approximately constant over the length of the atomic sample as the Rayleigh range is longer than the length of the cell.
For 100$^{\circ}$C vapour with most probable velocity $v_{\rm p}=215$~m/s and typical beam size $w_{\rm p}=w_{\rm c}=20$~$\mu$m, the most probable interaction time is $t_{\rm p}\approx90$~ns. This interaction time is comparable to the lifetime of both the intermediate state $\tau_{6P}=30.4$~ns and the excited state $\tau_{\rm 7S}=48.2$~ns.

First we consider the effect of applying only the probe laser on resonance $\Delta_{\rm p}/2 \pi=0$~MHz. Atoms with velocities around $v=0$~m/s are excited on the $F=4$ $\rightarrow$ $F'=5$ transition. As this is a closed transition, the atoms cannot enter the lower hyperfine level of the ground state and cycle on this transition until saturation is achieved. Due to the one-photon Doppler shift of the probe laser $\Delta_{\rm 1ph}=-k_{\rm p}v$, non-zero velocity classes around $v=213.85$~m/s and $v=385.10$~m/s are able to excite the $F=4$ $\rightarrow$ $F'=4$ and $F=4$ $\rightarrow$ $F'=3$ transitions respectively. As these transitions are open, atoms can enter the lower hyperfine level and become dark with respect to the probe laser. 

Now we consider the effect of applying both the probe and coupling laser simultaneously. The change in probe transmission as a function of coupling detuning is investigated for various beam waists and therefore atom-laser interaction times, as shown in Figure~\ref{fig:fig3}. When the coupling laser is on resonance $\Delta_{\rm c}/2 \pi=0$~MHz, a transparency feature is observed. This corresponds to the excitation of atoms with velocities around $v=0$~m/s on the $F=4$ $\rightarrow$ $F'=5$ $\rightarrow$ $F''=4$ two-photon transition. The probe transition is no longer closed and atoms can now decay to the lower hyperfine level of the ground state via hyperfine levels of the intermediate state. This process is known as Double Resonance Optical Pumping (DROP) \cite{Moon2004} and is a time-dependent transfer of atoms to the dark hyperfine state via a two-photon excitation process. 

For small beam sizes, and therefore short interaction times, the transparency feature at $\Delta_{\rm c}/2\pi=0$~MHz becomes absorptive. When the interaction time is sufficiently short, the atom does not remain in the beam long enough to decay from the excited  and intermediate state. As a result, the DROP transparency feature does not occur. Instead, in the very short interaction time domain $w_{\rm p}<30$~$\mu$m, the probe light is absorbed on resonance because atoms are being transferred to the upper state. 

However, for all interaction times, the feature at $\Delta_{\rm c}/2\pi=105$~MHz remains absorptive. This detuning corresponds to the two-photon detuning $\Delta_{\rm 2ph}=-(k_{\rm p}-k_{\rm c})v$ required to be resonant with the off resonant velocity class excited by the probe laser to the intermediate hyperfine level $F'=4$. When the coupling laser is applied to these open transitions, the decay routes to the lower hyperfine ground state are reduced. This is because atoms can now be excited through $F''=4$ to $F'=5$ which can only decay to the upper hyperfine ground state. As a result, the `openness' of the open transitions is reduced and the probe is absorbed more when the coupling laser is applied. 

\begin{figure}[t]
    \includegraphics[width=3.3in]{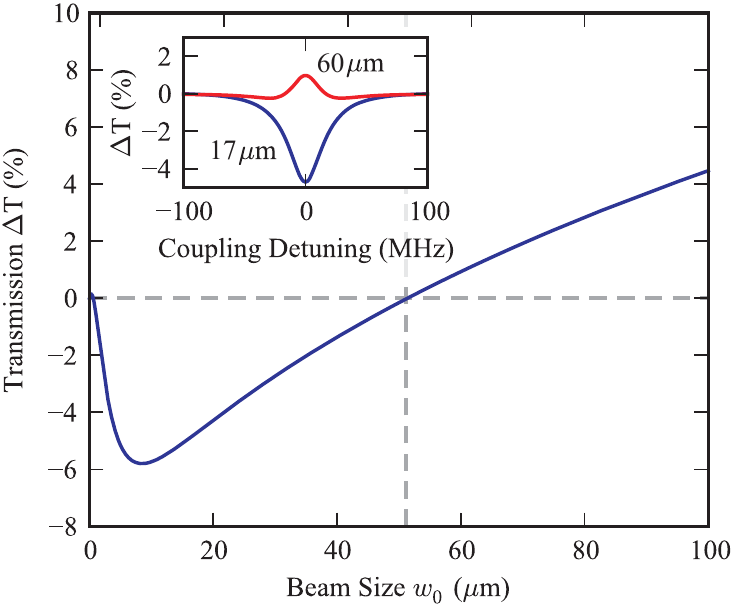}
    \caption{$6S-6P-7S$ theory: Change in probe transmission as a function of beam waist $w_{\rm 0}$ with $\lambda_{\rm p} = 852$~nm and $\lambda_{\rm c} = 1470$~nm when the coupling laser is on resonance. Inset: Change in probe transmission as a function of coupling laser detuning for two beam waists. Parameters: $\Omega_{\rm p}=2\pi \times8$~MHz, $\Omega_{\rm c}=2 \pi \times 22$~MHz, $a=0.6$ and $c=0.5$.}
    \label{fig:fig4}
\end{figure}

This interaction-time dependence of the transmission spectrum is fully reproduced by our model system, where the finite interaction time is included via the transit-time decay rate $\Gamma_{\rm tt}$. Figure~\ref{fig:fig4} shows the calculated probe transmission on resonance for the $F'=5$ intermediate state ($\Delta_{\rm c}/2\pi=0$~MHz), reproducing the change from absorptive for beam sizes $w_{0}<50$~$\mu$m to transmissive for larger beams. The exact position of the frontier between the two regimes in the simulation strongly depends on the Rabi frequencies. We attribute the discrepancy on the position of this frontier ($20$~$\mu$m experimentally versus $50$~$\mu$m theoretically) to the assumption of a constant Rabi frequency in the model, whereas in reality it is Gaussian-shaped. Our model also reproduces the behaviour of the other feature at $\Delta_{\rm c}/2\pi=105$~MHz ($F'=4$ intermediate state).

In this system EIT is still present but strongly reduced by the decay from the upper excited state and completely eclipsed by optical pumping effects. We find that there are two distinct time domains defined by the lifetimes of the intermediate and upper state. The response of the atomic system to the probe field is significantly different in these two regimes. For long interaction times optical pumping prevails, whereas for short interaction times, two-photon absorption dominates.

\section{Long-lived upper state}
\label{section:Stuttgart}

\begin{figure}[b]
    \includegraphics[width=3in]{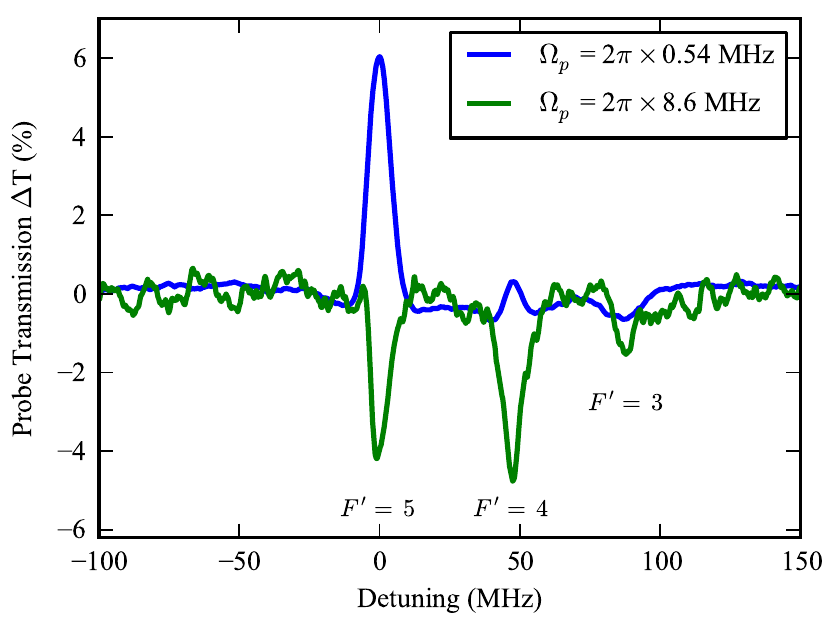}
    \caption{$6S-7P-32S$ experiment: Change in probe transmission for two different probe Rabi frequencies $\Omega_{\rm p} = 2\pi \times 0.54$~MHz (blue) and $\Omega_{\rm p} = 2\pi \times 8.6$~MHz (green) with $\lambda_{\rm p} = 455$~nm and $\lambda_{\rm c} = 1070$~nm. The coupling Rabi frequency is $\Omega_{\rm c}=2\pi \times 3.0$~MHz. The features at zero detuning corresponds to the intermediate state $|7P_{3/2}, F'=5 \rangle$, discussed here. The other two at $48$~MHz and $86$~MHz correspond to $F'=4$ and $F'=3$ respectively. Note that due to the measurement technique, no absolute transmission scale could be generated. Therefore the amplitude of the two traces are arbitrary and cannot be compared to one another.}
    \label{fig:fig5}
\end{figure}

\begin{figure}[t]
    \includegraphics[width=3in]{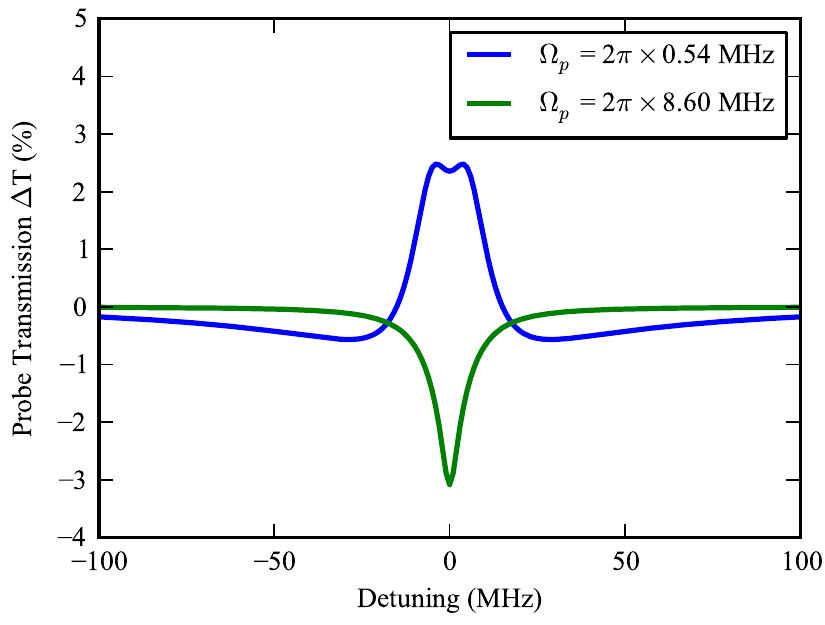}
    \caption{$6S-7P-32S$ Theory: Change in probe transmission for the same Rabi frequencies as in Figure~\ref{fig:fig5} for the $|7P_{3/2}, F'=5 \rangle$ intermediate state, with $\lambda_{\rm p} = 455$~nm and $\lambda_{\rm c} = 1070$~nm. The double structure around the resonance is a consequence of the Autler-Townes splitting by the probe laser. Parameters: $\Omega_{\rm p}=2\pi \times 0.54$~MHz and $2\pi \times 8.6$~MHz, $\Omega_{\rm c}=2\pi \times 3.0$~MHz, $w=1$~mm, $T=60^\circ$C, $a=0$, $b=0.55$ and $c=0.5$.}
    \label{fig:fig6}
\end{figure}

Next, we discuss the spectroscopy of a three-level ladder system with ``inverted''-wavelengths ($\lambda_p < \lambda_c$) where the upper level is a long-lived Rydberg state. The $|6S_{1/2}, F=4 \rangle \rightarrow |7P_{3/2}, F'=3,4,5 \rangle$ transition represents the probe transition and the $|7P_{3/2}, F'=3,4,5 \rangle \rightarrow |32S_{1/2} \rangle$ transition the coupling transition (see Figure~\ref{fig:fig1}). Such an excitation scheme is advantageous compared to standard non-inverted schemes (used for example in \cite{Mohapatra2008}) since the excitation to the Rydberg state at around $1070$~nm lies in the range of fibre lasers. The much higher output power, compared to diode lasers or Ti:Sa lasers, compensates for the small dipole matrix element and one can achieve coupling strengths comparable to the lower $6S_{1/2} \rightarrow 7P_{3/2}$ transition. 

Experimentally, both beams are linearly polarized with orthogonal polarizations and are overlapped in the vapour cell in a counter-propagating configuration. The $455$~nm probe beam is stabilized to the $|6S_{1/2}, F=4 \rangle \rightarrow |7P_{3/2}, F'=5 \rangle$ transition and the $1073$~nm coupling laser is scanned over the $|7P_{3/2}, F'=3,4,5 \rangle \rightarrow |32S_{1/2} \rangle$ transitions. As discussed before, the wavelength mismatch of our configuration dramatically reduces the visibility of the EIT peak. To observe a sufficiently large spectroscopy signal lock-in amplification was used. The coupling beam was intensity-modulated using an acousto-optic modulator (AOM) chopped at  $20$~kHz. Additionally, the demodulated probe signal was averaged over 32 traces. The cell had a length of $7.5$~cm and was heated to $60^\circ$C. Both beams were collimated with beam waists of $1.5$~mm for the $1073$~nm laser and $1.0$~mm for the $455$~nm laser. 

Two experimental probe transmission spectra are shown in Figure~\ref{fig:fig5}. Three structures are visible and can be identified with the three dipole-allowed hyperfine transitions $|6S_{1/2}, F=4 \rangle \rightarrow |7P_{3/2}, F'=3,4,5 \rangle$, where the lasers address three velocity classes fulfilling the two-photon resonance condition. A striking feature is the change of transparency to absorption by just changing the intensity of one of the two laser beams. In the following, we will focus on the excitation via the $F'=5$ intermediate state, but our results can be easily extended to the other allowed hyperfine intermediate states. 

\begin{figure}[b]
    \includegraphics[width=3in]{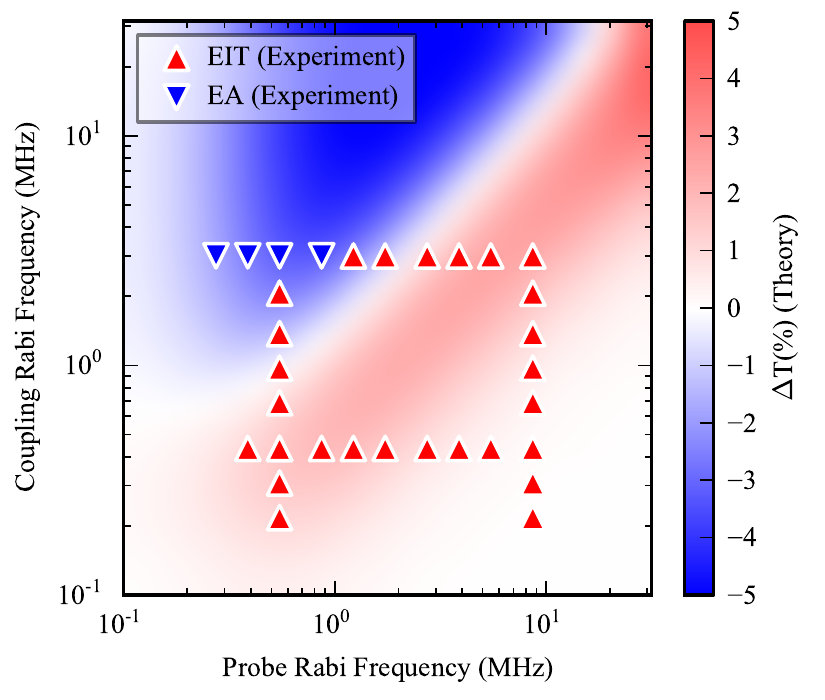}
    \caption{$6S-7P-32S$ theory: Foreground: qualitative representation of the experimental change in probe transmission on resonance. Background: theoretical change in probe transmission over the Rabi frequencies $\Omega_{\rm p}$ and $\Omega_{\rm c}$,  with $\lambda_{\rm p} = 455$~nm and $\lambda_{\rm c} = 1070$~nm. Parameters: $w=1$~mm, $T=60^\circ$C, $a=0$, $b=0.55$ and $c=0.5$.}
    \label{fig:fig7}
\end{figure}

These enhanced transmission (EIT) and enhanced absorption (EA) features are fully reproduced by our 5-level model. Simulated probe transmission traces with the same peak Rabi frequencies as in Figure~\ref{fig:fig5} are shown in Figure~\ref{fig:fig6}. The splitting in the blue curve around the resonance frequency has its origin in an Autler-Townes splitting of the intermediate state induced by the probe beam. The absence of the feature in the experimental spectra might be explained by the fact that the atoms cross the beam, therefore experiencing the Gaussian-distributed Rabi frequencies over the interaction time. The decay rates and branching ratios were matched to the real values, and the transit time effect was included for the corresponding temperature and beam size. A $100$~kHz laser induced dephasing was included, matching our laser linewidths on the timescale of our experiment. 
We found that including a laser induced dephasing was crucial in order to understand the experimental data. Indeed any decoherence effect reduces the EIT, which is already weak because of the wavelength ratio.  

We studied the dependence of the transmission signal on resonance (EIT or enhanced absorption) on both probe and coupling Rabi frequencies. The theoretical and experimental results are combined in Figure~\ref{fig:fig7}. The model system reproduces the experimentally observed transition from EIT to EA very well. We attribute the little discrepancy between theory and experiment to the fact that the exact position of this crossover in the simulation was found to be strongly dependent on various parameters, such as the laser linewidths, transit time and branching ratios. The experimental parameters were used as input parameters for the simulations in Figure~\ref{fig:fig7}.

Since the upper state is a long-lived Rydberg state with ${\rm 27 \mu s}$ lifetime, optical pumping effects from the upper state like DROP are negligible. The lower transition is however not closed, since the atoms can decay from the $|7P_{3/2} \rangle$ state to the $|7S_{1/2} \rangle$ and $|5D_{3/2, 5/2} \rangle$ states. 
From our simulations we observe that all of the following conditions have to be met, to reproduce the experimentally observed transition from EIT to EA: (i) the wavelengths are mismatched such that $\lambda_p < \lambda_c$, (ii) a laser induced dephasing is included, (iii) the first transition is open. Otherwise there exists no region of enhanced absorption and only EIT remains. 

Because of the ``inverted'' wavelength configuration, the EIT is a tiny feature, therefore extremely sensitive to other phenomena that are otherwise negligible. In particular for our system, having an open transition on the first excitation step acts as a decoherence on the system, simultaneously increasing the absorption and ``washing out'' coherent phenomena such as EIT. The finite laser linewidth further reduces the EIT signal, as well as it allows for more absorption around the resonance because of line broadening. 

\section{Conclusion}

To conclude, we have shown that EIT can be observed in ladder systems with ``inverted'' wavelengths. However, it is strongly affected by various incoherent effects due to additional decay channels and optical pumping, and EIT is very likely not to be the dominating effect. With our effective model based on five levels we were able to reproduce the experimentally observed features of two different physical implementations, which was not possible if only four states were included. Therefore we believe that this model will serve as a valuable design tool for all kinds of experiments involving ladder schemes. 

To date, very few studies have been performed involving ``inverted'' 3-level ladder schemes. Our results show that it is possible to obtain a good spectroscopic signal from these systems. These methods could be used for example as frequency references for further experiments, or as a spectroscopic tool.

\section*{Acknowledgments}

The authors would like to thank T.~Pfau, H.~K\"{u}bler, S.~Hofferberth, 
J.~Keaveney, U.~M.~Krohn and I.~G.~Hughes for fruitful discussions, as well as C.~Veit for experimental input.
This project was supported by the Carl-Zeiss-Stiftung and the UK Engineering and Physical Sciences Council. 
RL is indebted to the Baden-W\"{u}rttemberg Stiftung for the financial support of this research by the Eliteprogramme for Postdocs. RR acknowledges financial support from the
Landesgraduiertenf\"{o}rderung Baden–W\"{u}rttemberg. KJW acknowledges financial support from Durham University.

\bibliography{references}

\end{document}